# Simultaneous phase and amplitude aberration sensing with a liquid-crystal vector-Zernike phase mask

**David S. Doelman,**[1,*] **Fedde Fagginger Auer,**[1] **Michael J. Escuti,**[2] and **Frans Snik**[1]

[1]Leiden Observatory, Leiden University, P.O. Box 9513, 2300 RA Leiden, The Netherlands
[2]Department of Electrical and Computer Engineering, North Carolina State University, Raleigh, North Carolina 27695, USA
*Corresponding author: doelman@strw.leidenuniv.nl

**We present an enhanced version of the Zernike wavefront sensor (WFS), which simultaneously measures phase and amplitude aberrations. The "vector-Zernike" WFS consists of a patterned liquid-crystal mask, which imposes a $\pm\pi/2$ phase on the point spread function core through the achromatic geometric phase acting with the opposite sign on opposite circular polarizations. After splitting circular polarization, the ensuing pupil intensity images are used to reconstruct the phase and the amplitude of the incoming wavefront. We demonstrate reconstruction of the complex wavefront with monochromatic lab measurements and show in simulation the high accuracy and sensitivity over a bandwidth up to 100%.** © 2018 Optical Society of America

https://doi.org/10.1364/OL.44.000017

Provided under the terms of the OSA Open Access Publishing Agreement

The introduction of phase-contrast microscopy by Zernike [1–4] was a revolution for the field of biological sciences. While a standard microscope measures intensity variations due to absorption, a phase-contrast microscope measures the phase differences from variations in the index of refraction by coupling the phase differences into intensity variations with a phase mask inside the microscope. Otherwise, unseen transparent structures could now be imaged in great detail. Another application of the phase-contrast method is the Zernike wavefront sensor (ZWFS). The ZWFS has a similar phase mask with a dot that offsets the phase of the core of the point spread function (PSF) by $\pi/2$. The optimal size of the dot is $1.06\ \lambda/D$, where $D$ is the pupil diameter, and $\lambda$ is the wavelength. The phase-shifted PSF core interferes with the rest of the PSF, and phase aberrations are converted to electric field amplitude variations in the subsequent pupil plane [4–6]. Thus, a phase-only aberration can be measured directly with one intensity measurement. Extremely sensitive systems such as direct-imaging instruments for exoplanet detection use the ZWFS to minimize instrumental aberrations. Non-common path correction with the ZWFS has been successfully demonstrated on-sky [7], and picometer precision has been achieved in the lab for the wide-field infrared survey telescope [8].

One major advantage of the ZWFS is that it is the most photon-efficient WFS with a sensitivity factor for photon noise of unity [6,9]. In addition, the ZWFS constitutes a simple implementation for any optical system that only requires a focal plane mask and the ability to image the pupil plane. However, the dynamic range is limited, and the mask applies (wavelength-dependent) scalar phase shifts. In addition, a reference pupil image is needed for static amplitude aberration correction. Therefore, time-variable amplitude aberrations are reconstructed as phase aberrations. Static and dynamic amplitude variations can come from the contamination of optics, degradation in optical performance, Fresnel propagation effects, and scintillation. Several systems that measure both amplitude and phase have been proposed; however, these systems are complicated [10,11] or require diversity in time such as the phase-shifting ZWFS, where a variable-phase Zernike mask is used to reconstruct the complex wavefront with four consecutive measurements [12].

In this Letter, we present a liquid-crystal version of the ZWFS, the "vector-Zernike" wavefront sensor (vZWFS) that enables the simultaneous measurement of both phase and amplitude aberrations. This simple upgrade only requires the replacement of the Zernike mask, in addition to splitting circular polarization. The property of optimal photon efficiency is maintained by the vZWFS, and a liquid-crystal focal plane mask applies an achromatic geometric phase, yielding an improved broadband performance [13]. The geometric phase applied by the mask, $\Delta\theta$, is independent of the wavelength and only depends on the fast-axis orientation, $\alpha$, of the mask and the handedness of the circular polarization state of the incoming light [14]:

$$\Delta\theta(x, y) = \pm 2\alpha(x, y). \quad (1)$$

Writing complex fast-axis orientation patterns is enabled by liquid-crystal direct-write technology [15], and achromatizing the half-wave retardance is enabled by stacking self-aligning liquid-crystal layers to form a multi-twist retarder [16]. The precise writing capability and broadband performance of this technology have been verified in the lab for other optical elements [14,17]. A Zernike mask manufactured with liquid-crystal technology simultaneously applies $+\pi/2$ phase to the right-handed circularly polarized fraction of the PSF core and $-\pi/2$ to the left-handed circularly polarized fraction of the PSF core. We split the light with opposite handedness with a quarter-wave plate (QWP) and a polarizing beam splitter (PBS), as shown in Fig. 1.

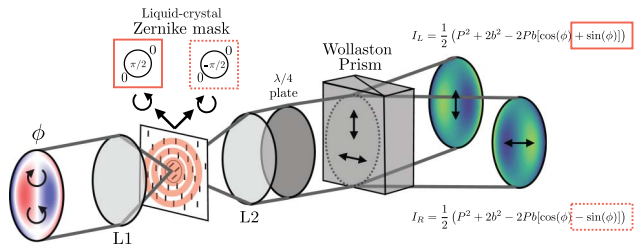

**Fig. 1.** Layout of the vZWFS. The intensity distribution of the pupils depends on the incoming phase and sign of the polarization state.

To describe the resulting intensity in the two pupils, we adopt the formalism from N'Diaye et al. [5]. The electric field at the entrance pupil, $\Psi_A$, is defined as

$$\Psi_A = Pe^{i\phi} = P_0(1-\epsilon)e^{i\phi}. \quad (2)$$

Here $P$ is the normalized amplitude such that $P_0$ is the pupil function. The variables $\epsilon = \epsilon(u,v)$ and $\phi = \phi(u,v)$, respectively, are the amplitude aberration and phase aberrations for each position in the pupil plane. Following the derivation of N'Diaye et al. [5] and assuming $\theta = \pm\pi/2$, the intensity of the two pupils ($I_L$ and $I_R$) can be written as

$$I_L = \frac{1}{2}(P^2 + 2b^2 - 2Pb[\cos(\phi) + \sin(\phi)]), \quad (3)$$

$$I_R = \frac{1}{2}(P^2 + 2b^2 - 2Pb[\cos(\phi) - \sin(\phi)]), \quad (4)$$

where $b$ is the convolution of the Fourier transform of the binary Zernike mask and the pupil plane electric field. If the aberrations are small, $b$ can be approximated by $b_0$, using the pupil function instead of the pupil plane electric field ($\Psi_A = P_0$).

Splitting opposite circular polarization states introduces the factor by one-half and creates two pupils with different intensities, depending on the sign of $\sin(\phi)$. This is used to measure both the amplitude and phase aberrations in a similar fashion as Wallace et al. [12]. They use the sum and the difference of four intensity measurements to directly calculate the amplitude and phase aberrations. Here the phase is measured in the same way, except that the amplitude is measured as the square root of the pupil intensity. The sum and the difference of the pupil intensities are given by

$$I_R + I_L = P^2 + 2b^2 - 2Pb\cos(\phi), \quad (5)$$

$$I_R - I_L = 2Pb\sin(\phi). \quad (6)$$

Using these equations and the identity $\cos^2(\phi) + \sin^2(\phi) = 1$, we solve for the amplitude $P$ and the phase $\phi$. The exact reconstruction is given by

$$P = \sqrt{I_R + I_L + \sqrt{4b^2(I_R + I_L) - (I_R - I_L)^2 - 4b^4}}, \quad (7)$$

$$\phi = \arcsin\left(\frac{I_R - I_L}{2Pb}\right). \quad (8)$$

Solving these equations can become numerically unstable for real wavefront sensor applications. Similar to N'Diaye et al. [5], we assume that the vZWFS operates in the low-aberration regime, where $\phi$ dominates the reconstruction. Therefore, $b \approx b_0$, and, for the phase reconstruction, $P = 1$. Then we calculate $P$ using the approximated values for $\phi$:

$$\phi \approx \frac{(I_R - I_L)}{2b_0}, \quad (9)$$

$$P \approx \sqrt{I_R + I_L - b_0^2(2 - \cos^2(\phi))} - b_0\cos(\phi). \quad (10)$$

Measuring both $P$ and $\phi$ allows us to estimate the aberrated electric field and approximate $b$. This process is iterated and converges in the linear regime to the right $b$, $P$, and $\phi$. Note that the reconstruction algorithms do not require many complex calculations such as matrix multiplications to recover the complex wavefront. Therefore, the reconstruction is suitable for very fast wavefront reconstruction.

To compare the effect of amplitude aberrations in the reconstruction for both the ZWFS and the vZWFS, we perform numerical simulations of both systems with the Python package HCIPy [18]. We start with an unobstructed pupil with both phase and amplitude aberrations, as shown in the left column of Fig. 2. The input phase aberration has a $\sigma_\phi = 0.22$ rad rms and 1.17 rad peak-to-valley (PtV). The normalized amplitude aberration has a $\sigma_\epsilon = 0.6\%$ rms ($\sigma_I = 1.3\%$) and 3.7% PtV (I: 7.2% PtV), where I is the intensity. We reconstruct the phase for the ZWFS using Eq. (14) in N'Diaye et al. [5]. For the vZWFS, we use three different methods to calculate the phase and the amplitude. The first method (*Linear*) uses modal-based wavefront reconstruction to reconstruct 200 Zernike mode coefficients of the phase, and the amplitude uses Eq. (10). The second method (*second order*) uses

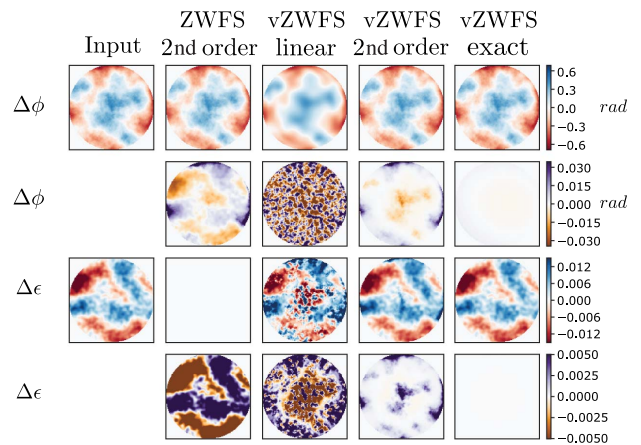

**Fig. 2.** Comparison in the simulated reconstruction of both phase and amplitude between the ZWFS and the vZWFS. The phase aberration is shown in the top row, followed by the residual phase, i.e., the difference between the reconstructed and the input phase. The two bottom rows contain the same with amplitude aberrations. The different columns show the results for different measurement and reconstruction methods.

Eqs. (9) and (10) to reconstruct the aberrations directly from the intensity measurements with five iterations. The third method (*Exact*) uses Eqs. (7) and (8) using $b = b_0$ and five iterations. We compare the standard deviation of the residual phase between the ZWFS and the vZWFS, and find that the nonlinear algorithms of the vZWFS outperform the ZWFS. For the second row in Fig. 2, from left to right, starting from the input, $\sigma_{\Delta\phi} = [0.221, 0.013, 0.029, 0, 007, 0.001]$ rad rms. The phase reconstruction of the classical ZWFS in the second column of Fig. 2 shows that the residual phase aberration in the pupil is caused by the amplitude aberration that it cannot directly correct for. This is different from all three vZWFS reconstruction methods, where the residual phase is dominated by the inability to reconstruct high-frequency aberrations with the first 200 Zernike modes (*Linear*), approximations used in the phase calculation (*2nd order*), and the dynamic range of the vZWFS (*Exact*). No significant influence of the amplitude aberration is seen in the residual phase. The amplitude aberration measurements of the vZWFS are dominated by phase estimation errors. The residual amplitude aberrations from left to right, starting from the input, are $\sigma_{\Delta\epsilon} = [1.3, 1.3, 1.4, 0.4, 0.02]\%$ rms.

For practical implementations, we look into the influence of imperfect optics shown in Fig. 1 and noise on the performance of the vZWFS. Measuring two pupils increases the readout noise by a factor of 2, and combining the two pupil measurements suffers from flat-fielding effects. In addition, the vZWFS setup requires the use of a liquid-crystal Zernike mask, a QWP and a PBS. The vector-Zernike mask has an efficiency (the amount of light that acquires the geometrical phase) that depends on the retardance of the liquid-crystal film [16]. When the phase mask has an offset of the half-wave retardance, $\Delta\delta_{HW}$, the incoming unpolarized light that does not acquire the phase is split by the PBS and adds a background proportional to the pupil intensity. Similar contaminations happen with retardance deviations of the QWP and the rotation of the PBS. The resulting left pupil intensity distributions (identical to the right pupil) from an imperfect half-wave plate, QWP, and PBS, respectively, are given by

$$I_{L,\Delta 1} = I_L \cos^2\left(\frac{1}{2}\Delta\delta_{HW}\right) + \frac{1}{2}P^2 \sin^2\left(\frac{1}{2}\Delta\delta_{HW}\right), \quad (11)$$

$$I_{L,\Delta 2} = I_L \cos^2\left(\frac{1}{2}\Delta\delta_{QW}\right) + \frac{1}{2}(I_L + I_R)\sin^2\left(\frac{1}{2}\Delta\delta_{QW}\right), \quad (12)$$

$$I_{L,\Delta 3} = I_L \cos^2(\Delta\theta) + I_R \sin^2(\Delta\theta). \quad (13)$$

To the first order, the difference of the pupil intensities is most affected by the extra terms. All terms proportional to $\sin^2()$ drop out, and the difference now is proportional to a $\cos^2()$ term that reduces the response. The reduced response can be corrected with a gain factor from calibration. The sum of the pupil intensities is only affected by the deviation from half-wave retardance of the vector-Zernike mask; all other terms sum to the total intensity $I_L + I_R$. The chromatic retardance can be controlled with liquid-crystal technology such that $\Delta\delta_{HW} < 16°$ [16] for a large bandwidth, i.e., the leakage is <2% of the intensity.

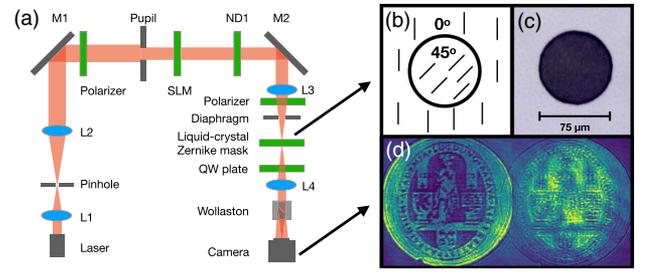

**Fig. 3.** (a) Layout of the vZWFS setup. We generate a clean beam with a laser (633 nm) and a pinhole. A SLM is operated in the phase-mostly configuration with two polarizers and is used to generate phase aberrations with a complicated pattern (i.e., the Leiden University logo). This configuration also generates some amplitude aberrations. The light is focused on the vector-Zernike mask with a spot diameter of 75 μm, corresponding to ~1λ/D. The liquid-crystal orientation is shown in panel (b), and a parallel polarizer microscopic image is shown in panel (c). The detector image with aberrated pupils is shown in panel (d), showing some pupil overlap because of the splitting angle of the Wollaston.

As a proof of principle, we demonstrate the vZWFS with a monochromatic setup. We manufactured a liquid-crystal vector-Zernike mask with a central spot diameter of 75 μm, printed on a glass substrate (1 in diameter, 1 mm thick, BK7) with an effective pixel size of 2 μm. The retardance was tuned to be λ/2 at 633 nm with a single liquid-crystal layer. Figure 3(b) shows the fabricated mask that does not contain any significant defects. The edges have a smooth fast-axis transition region from 0° to 45° of 2–3 micron. This can be mitigated by modifying $b$ for calculating $P$ and $\phi$, although the effect is negligible. We test the performance of the vector-Zernike using the setup shown in panel (a) of Fig. 3. We use a LC2002 transmissive spatial light modulator (SLM) from HoloEye in the phase-mostly configuration to generate phase aberrations. In this configuration, the rotation of the liquid-crystals in combination with the polarizers also generates amplitude aberrations. The Wollaston prism is a WPQ10 prism from Thorlabs, and the QWP is an achromat (AQWP05M-600) from Thorlabs. Panel (d) in Fig. 3 shows a raw image after applying a binary logo of Leiden University on the SLM. For the characterization of phase and amplitude reconstruction, we minimize the influence of the system by combining measurements with positive and negative phase applied by the SLM. We mean-combine them to remove the non-SLM aberrations, and we bin the image per 2 pixels to increase the signal to noise. We also take images without any phase applied on the SLM and use them for calculating the pupil intensity necessary for normalizing the aberrated images without taking out the vector-Zernike mask. Note that taking the average pupil intensity does not normalize the measurements correctly. Assuming $\phi \ll 1$, we solve Eq. (5) with $\cos\phi = 1$, such that $P$ is given by

$$P = \sqrt{(I_L + I_R)/(1 + 2b_0^2/P^2 - 2b_0/P)}. \quad (14)$$

When $\epsilon \ll 1$, the normalized quantity $b_0/P \approx b_0/P_0$, and we calculate the normalization factor of the pupil images by estimating $b_0$ and averaging $P$ over the pupil. We calculate the phase and amplitude from the two aberrated normalized pupils with Eqs. (9) and (10) and are shown in Fig. 4. Both measurements contain Fresnel diffraction effects from the out-of-plane diaphragm used as a field stop. Determining the accuracy of the reconstruction is not possible, as we do not know the details of this SLM.

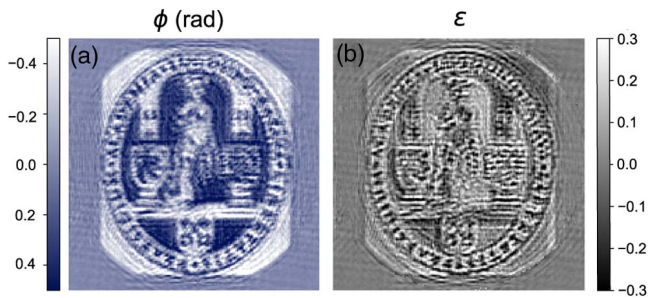

**Fig. 4.** (a) Reconstructed phase and (b) reconstructed amplitude aberration using the vZWFS.

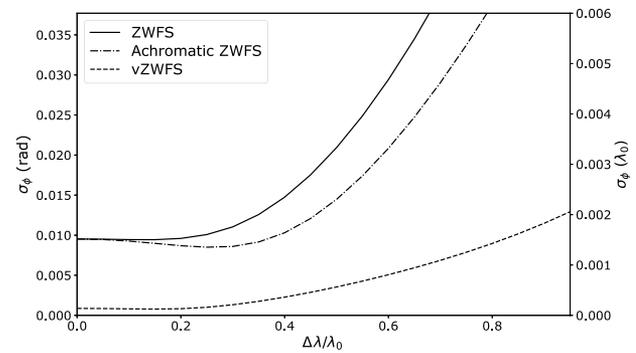

**Fig. 5.** Residual wavefront aberrations as a function of the bandwidth for both the classical ZWFS and the vZWFS. The "achromatic ZWFS" uses the classical ZWFS reconstruction on one pupil of the vZWFS.

One additional advantage of the liquid-crystal vZWFS is that the phase is applied achromatically; see Eq. (1). The classical ZWFS has decreased performance with increasing spectral bandwidth because of the chromatic phase of the focal plane mask, in combination with the PSF scaling with wavelength. Both contribute to a suboptimal coupling of phase aberrations in the system to intensity variations in the ZWFS pupil plane [5]. We study the effect of the achromatic phase and the different reconstruction on the broadband performance of the vZWFS in simulation, taking into account the spatial scaling of the PSF with wavelength. For this, we exclude amplitude aberrations. Figure 5 shows the results for both wavefront sensors up to 100% bandwidth, i.e., $\lambda/\lambda_0 = 0.5 - 1.5$. We assume a $1/\lambda$ scaling for the applied phase of the classical Zernike mask centered around $\lambda_0$, and $b_0$ is calculated at $\lambda_0$ as well. We investigate the influence of only the achromatic phase by using the classical ZWFS reconstruction on one of the vZWFS pupils, the "achromatic ZWFS." Note that all other chromatic effects, including leakage from polarization optics, are not taken into account in this simulation. Figure 5 shows that the vZWFS with exact reconstruction outperforms the classical ZWFS for all bandwidths. Similarly, the classical reconstruction with the "achromatic ZWFS" improves the broadband performance, in agreement with Bloemhof [13]. The gain of using only achromatic phase is not as significant as the gain from the reconstruction using two pupils. The reason is that the classical ZWFS uses a second-order reconstruction that is sensitive to the chromatic $b$, while the vZWFS allows for an exact solution. Iterating the estimate of $b$ with the updated phase and amplitude estimates removes the bias introduced by the chromaticity. Overall, we show in simulation that the vZWFS can handle an increased bandwidth, up to 100%.

We conclude that the liquid-crystal vZWFS is a simple yet powerful upgrade of the classical ZWFS. We show that both amplitude and phase aberrations can be measured simultaneously by replacing the Zernike mask with a liquid-crystal version and splitting circular polarizations in two pupils. We demonstrate both in simulations and with a monochromatic lab setup an improved reconstruction of the complex wavefront with the vZWFS. Like the ZWFS, the vZWFS is the most photon-efficient WFS [6], and we demonstrate with simulations the accurate reconstruction for bandwidths up to 100%. For direct imaging of exoplanets the vZWFS can be installed a parallel WFS beam path, alternated with a coronagraph [7], or built into a coronagraph [8].

**Funding.** H2020 European Research Council (ERC) (678194, FALCONER).